\documentclass[aip, apl, reprint, amsmath, graphicx, longbibliography]{revtex4-1}

\usepackage{graphics}
\usepackage{graphicx}
\usepackage{bm}
\usepackage{color}

\begin{document}


\title{Observation of quantum interference conductance fluctuations in metal rings with strong spin-orbit coupling}



\author{R. Ramos}
\email[]{present address: r.ramos@usc.es}
\affiliation{WPI-Advanced Institute for Materials Research, Tohoku University, Sendai 980-8577, Japan}
\affiliation{Centro de Investigaci\'{o}n en Qu\'{i}mica Biol\'{o}xica e Materiais Moleculares (CIQUS), Departamento de Qu\'{i}mica-F\'{i}sica, Universidade de Santiago de Compostela, Santiago de Compostela 15782, Spain}

\author{T. Makiuchi}
\affiliation{Department of Applied Physics, The University of Tokyo, Tokyo 113-8656, Japan}

\author{T. Kikkawa}
\affiliation{WPI-Advanced Institute for Materials Research, Tohoku University, Sendai 980-8577, Japan}
\affiliation{Department of Applied Physics, The University of Tokyo, Tokyo 113-8656, Japan}
\affiliation{Institute for Materials Research, Tohoku University, Sendai 980-8577, Japan}

\author{S. Daimon}
\affiliation{Department of Applied Physics, The University of Tokyo, Tokyo 113-8656, Japan}

\author{K. Oyanagi}
\affiliation{Faculty of Science and Engineering, Iwate University, Morioka 020-8551, Japan}

\author{E. Saitoh}
\affiliation{WPI-Advanced Institute for Materials Research, Tohoku University, Sendai 980-8577, Japan}
\affiliation{Department of Applied Physics, The University of Tokyo, Tokyo 113-8656, Japan}
\affiliation{Institute for Materials Research, Tohoku University, Sendai 980-8577, Japan}
\affiliation{Center for Spintronics Research Network, Tohoku University, Sendai 980-8577, Japan}
\affiliation{Advanced Science Research Center, Japan Atomic Energy Agency, Tokai 319-1195, Japan}


\date{\today}

\begin{abstract}
We investigated the magnetotransport properties of mesoscopic platinum nanostructures (wires and rings) with sub-100 nm lateral dimensions at very low temperatures. Despite the strong spin-orbit interaction in platinum, oscillations of the conductance as a function of the external magnetic field due to quantum interference effects was found to appear.  The oscillation was decomposed into Aharonov--Bohm periodic oscillations and aperiodic fluctuations of the conductance due to a magnetic flux piercing the loop of the ring and the metal wires forming the nanostructures, respectively. We also investigated the magnetotransport under different bias currents to explore the interplay between electron phase coherence and spin accumulation effects in strong spin-orbit conductors.  
\\
\end{abstract}


\maketitle 
Conductive materials with strong spin-orbit coupling are at the central stage of spintronics research. The mutual interaction between the charge and spin currents by the spin Hall effect (SHE) and its Onsager reciprocal: the inverse spin Hall effect (ISHE), has vastly expanded the field; enabling the investigation of spin-information transport,\cite{Kajiwara2010, Uchida2010i, Uchida2010, Wu2015, Seki2015, Wu2016, Ramos_2018, Lebrun2018, Oyanagi2019, Ramos2019} spin-orientation readout \cite{Nakayama2013, Hoogeboom2017, Hou2017, Fischer2018, Baldrati2018, Oyanagi2020} and electrical manipulation of the spin state of magnetic insulators.\cite{Avci2017, Moriyama2018, Chen2018} These phenomena are characterized by lengthscales in the nanometer range, therefore investigation of the fundamental transport properties at the nanoscale level can further expand the field, merging spin dependent transport phenomena and mesoscopic physics, and possibly leading to new functionalities of potential interest for future quantum technologies.\cite{Niita1999, Shun-Qing2004, Souma2005}\\
\begin{figure}[htp!] \center
    \includegraphics[width=8.5cm, keepaspectratio]{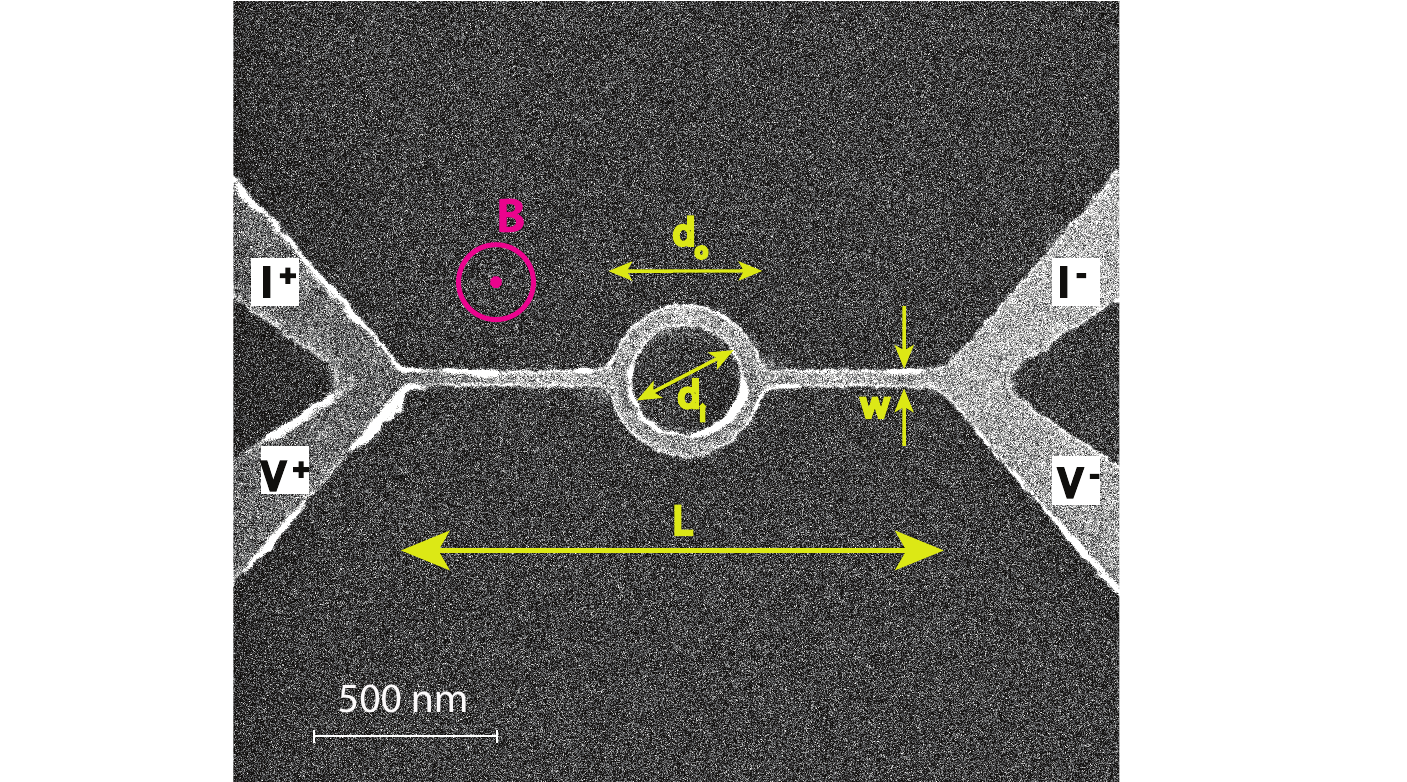}
  \caption{(Color online) Scanning electron microscope (SEM) image of a typical Pt ring (diameter 350 nm) fabricated for the observation of the Aharonov--Bohm conductance oscillations. The experimental geometry and sample dimensions are also schematically depicted.} \label{fig1}
\end{figure}
One interesting possibility is the study of quantum phase interference effects on the magnetotransport. These phenomena typically manifests in different ways: (1) a magnetoresistance background due to weak (anti-) localization effects\cite{Hikami1980, Bergmann1984} in conductive samples, (2) the appearance of a randomly oscillating conductance pattern, known as universal conductance fluctuations (UCF),\cite{Umbach1984, Lee1985, Lee1987} characteristic of the microscopy impurity configuration and resulting from a magnetic flux piercing the body of the sample in laterally confined structures (nanowires) and (3) the periodic Aharonov--Bohm (AB) oscillations \cite{Aharonov1959} superimposed to the aperiodic UCF background in nanoring structures with characteristic dimensions of the order of the electron phase coherence length.\\ 
 \begin{figure*}[htp!] \center
    \includegraphics[width=15cm, keepaspectratio]{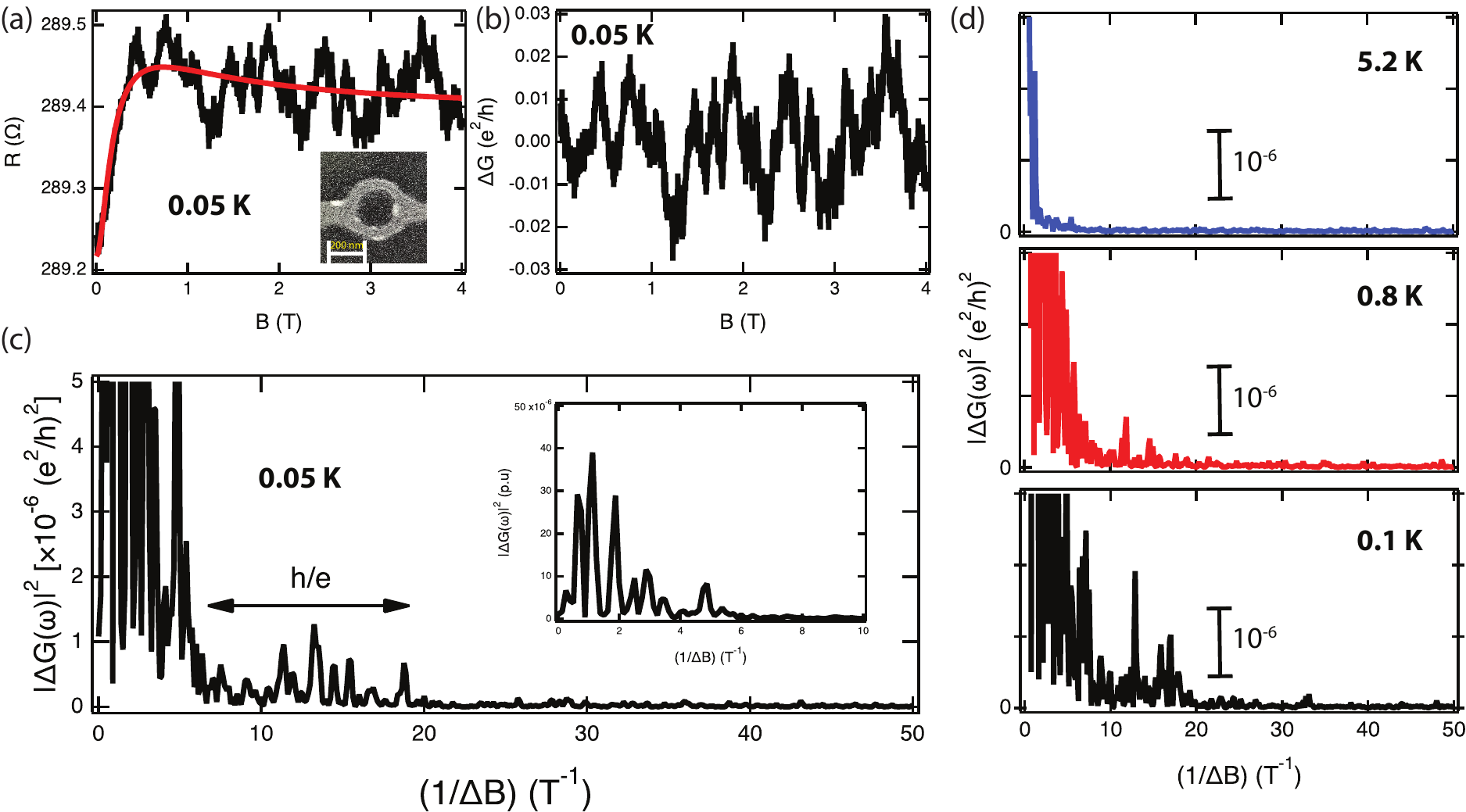}
  \caption{(Color online) (a) Longitudinal magnetoresistance measured in a Pt ring with an average diameter of 250 nm (see inset for an SEM image of the ring) measured at 0.05 K and the fitting (red line) of the magnetoresistance to the expression for the WAL of quasi-1D structures shown in Eq. \ref{eq1} . (b) Sample conductance ($\Delta G = \Delta R/R_0^2$), $R_0$ is the zero field resistance) estimated after removal of the WAL component ($\Delta R = R - R_\text{WAL}$). (c) Power spectrum obtained from the Fourier transform of the autocorrelation function of the magneto-conductance, showing clear peaks corresponding to periodic Aharonov--Bohm oscillations with periodicity corresponding to a flux of $h/e$ traversing the rings. The inset shows the detail of the power spectrum corresponding to the aperiodic universal conductance fluctuations (UCF). (d) Temperature dependence of the Fourier spectrum for (bottom to top) 0.1, 0.8 and 5.2 K.}
  \label{fig2}
\end{figure*}
The magnetoconductance oscillations have been previously studied in non-magnetic metal rings with coherence lengths in the micrometer range,\cite{Webb1985, Washburn1986, Milliken1987, Haussler2001} or metallic ferromagnets with relatively large coherence lengths.\cite{Kasai2002} The investigation of quantum phase interference effects in strong spin-orbit metals can further expand the capabilities of these materials for spintronics and possibly lead to new functionalities, as in the previously proposed spin interference devices without magnetic elements.\cite{Niita1999, Shun-Qing2004, Souma2005} However, the observation of conductance oscillation effects in metals with strong spin-orbit coupling and short phase coherence length has remained elusive, due to the difficulty to fabricate suitable nanostructures and the expected suppression of the electron phase coherence by the strong spin-orbit interaction.\\
\indent
Here we report the observation of magnetoconductance oscillations in Pt, a material with strong spin-orbit interaction and short phase coherence length. The observation of this effect in Pt opens the possibility to explore the mutual interaction between effects governed by the electron phase coherence and the spin degree of freedom, such as possible interactions between conductance oscillations and SHE-driven spin accumulation.\\
\indent
The samples investigated in the present study were fabricated on a 10 nm Pt film deposited at room temperature on a SiO$_2$/Si substrate by dc-magnetron sputtering (ULVAC QAM-4 STS). Two types of lateral nanostructures were fabricated: wires and rings, by combination of electron beam lithography and argon milling. In this work we investigate one nanowire and three nanorings of different diameters, with nominal dimensions: lateral width ($w$ = 40 nm), distance between current/voltage leads ($L$ = 1.5 $\mu$m) and diameters ($d$) of 250, 350 and 500 nm. Figure \ref{fig1} shows a typical scanning electron microscopy (SEM) image of one of the fabricated nanorings. The magnetotransport properties of the samples at mK temperatures were investigated in a $^3$He--$^4$He dilution refrigerator (KelvinoxMX200). The magnetoresistance was measured by a four probe method (13.1 Hz, 100 nA) with an ac resistance bridge (Lake Shore 372). The sweep rate of the magnetic field was kept at 12 mT/min to minimize heating effects by eddy currents of the sample holder.\\
\indent
Figure \ref{fig2} shows the longitudinal resistance of a Pt ring with an average diameter of 250 nm measured at 0.05 K. The data shows the typical positive magnetoresistance at low field due to weak anti-localization (WAL), characteristic of materials with a strong spin-orbit coupling and in agreement with previous reports in Pt wires and thin films.\cite{Niimi2013, Shiomi2014, Ryu2016} To determine the phase coherence length ($L_\phi$) and spin-orbit length ($L_\text{SO}$) we employ the Hikami--Larkin--Nagaoka formula for a 1D wire, valid when the width of the wires forming the ring structures is small compared to the electron coherence length:\cite{akkermans_montambaux_2007, Niimi2013} 
\begin{equation}
\label{eq1}
\frac{\Delta R_\text{WAL}}{R_\infty}=\frac{1}{\pi L}\frac{R_\infty}{\hbar/e^2}\left[\frac{3/2}{\sqrt{\frac{1}{L_\phi^2}+\frac{4}{3}\frac{1}{L_\text{SO}^2}+\frac{1}{3}\frac{w^2}{l_\text{B}^4}}}-\frac{1/2}{\sqrt{\frac{1}{L_\phi^2}+\frac{1}{3}\frac{w^2}{l_\text{B}^4}}}\right],
\end{equation}
where $\Delta R_\text{WAL} = R_\text{WAL} - R_\infty$, $R_\infty$, $L$ and $w$ are the weak localization correction factor, the resistance at high enough field, the length between contacts and width of the Pt nanostructures, and $e$, $\hbar$ and $l_\text{B} = (\hbar/eB)^{0.5}$ are the electron charge, the reduced Planck constant and the magnetic length, respectively. From the WAL fitting we can estimate the electron coherence and spin-orbit lengths, obtaining $L_\phi$ = 150 nm and $L_\text{SO}$ = 27 nm at 0.05 K. The magnetoconductance ($\Delta G$) estimated after subtraction of the WAL magnetoresistance background, and expressed in units of the conductance quantum ($e^2/h$) is shown in Figure \ref{fig2}b, we can see a clear fluctuation of the sample conductance characteristic of the electron phase interference effects. Before calculating the Fourier transform to estimate the spectrum of the oscillations frequencies, we calculated the autocorrelation function of $\Delta G$ to further reduce the random noise. Moreover, following this procedure we can directly obtain the power spectrum of the measured signal from the Fourier transform of the autocorrelation function.\cite{Landau2015} The obtained Fourier spectrum comprises two regions: a low frequency region (see inset of Fig. \ref{fig2}c) originated from the aperiodic fluctuations due to UCF\cite{Lee1985, Lee1987}, arising from interference effects within the body of the Pt and resulting in an aperiodic fluctuation pattern dependent on the defect distribution within the sample. At higher frequencies (Fig. \ref{fig2}c), we can distinguish the presence of Aharonov--Bohm oscillations due to the interference between electrons traveling clockwise and counterclockwise through the arms of the ring, which can be confirmed by the presence of peaks in the Fourier spectrum in the frequency range from 6 to 18 T$^{-1}$. These correspond to conductance oscillations due to a quantum of magnetic flux ($\phi_0 = h/e$) being enclosed within the area of the Pt rings (the frequency range for the AB oscillation in the Fourier spectrum is a result of the finite width of the Pt wires forming the ring structure). The presence of AB oscillations in Pt ring is quite surprising, given the short value of the estimated coherence length from the WAL fitting, being much shorter than the half-perimeter of the ring (390 nm). From the amplitude of the AB oscillations we can also obtain an estimate of the phase coherence length. For a fully coherent sample, the expected amplitude of the AB oscillations is 0.3 $e^2/h$. Whereas for a sample with a length of the ring arms larger than $L_\phi$, the amplitude of the oscillations has an exponential suppression factor of (2$L_\phi/U)\exp(-U/L_\phi)$, where U is the circunference of the ring.\cite{Haussler2001} Considering this expression and the amplitude of the AB oscillations in our measurement ($\sim$ 0.02 e$^2$/h at 0.05 K, Fig. \ref{fig2}b) we can estimate a coherence length of $L_\phi \sim$ 300 nm, within the same order of magnitude as the one obtained from the fit to the WAL expression. We can also obtain an estimate of the phase coherence length from the UCF measurement of a Pt wire, using the expression $L_\phi = \phi_0/(wB_\text{C, UCF})$\cite{Haussler2001}, where $B_\text{C,UCF}$ is the correlation field of the UCF, obtained from the conductance autocorrelation function.\cite{Lee1987, Alagha2010, Meng2019} Using this expression, we can estimate for a Pt nanowire a value of $B_\text{C,UCF}$ = 0.4 T (see Supplementary for estimation of  $B_\text{C, UCF}$) and a phase coherence length of $L_\phi$ = 260 nm, in reasonable agreement with the previously obtained values from the WAL fit and AB oscillation amplitude. The temperature dependence of the conductance oscillations is depicted in Fig. \ref{fig2}d, which shows the Fourier spectrum obtained for three different temperatures, we can observe that the amplitude of the AB oscillations decreases upon a temperature increase, as expected.\cite{Umbach1984, Webb1985, Washburn1986} Our result suggests that we can observe magnetoconductance oscillations due to electon phase interference effects, despite the much shorter phase-coherence length of Pt compared to previously studied materials.\cite{Webb1985, Washburn1986, Milliken1987, Haussler2001, Kasai2002} 
\begin{figure}[htp!] \center
    \includegraphics[width=8.5cm,keepaspectratio]{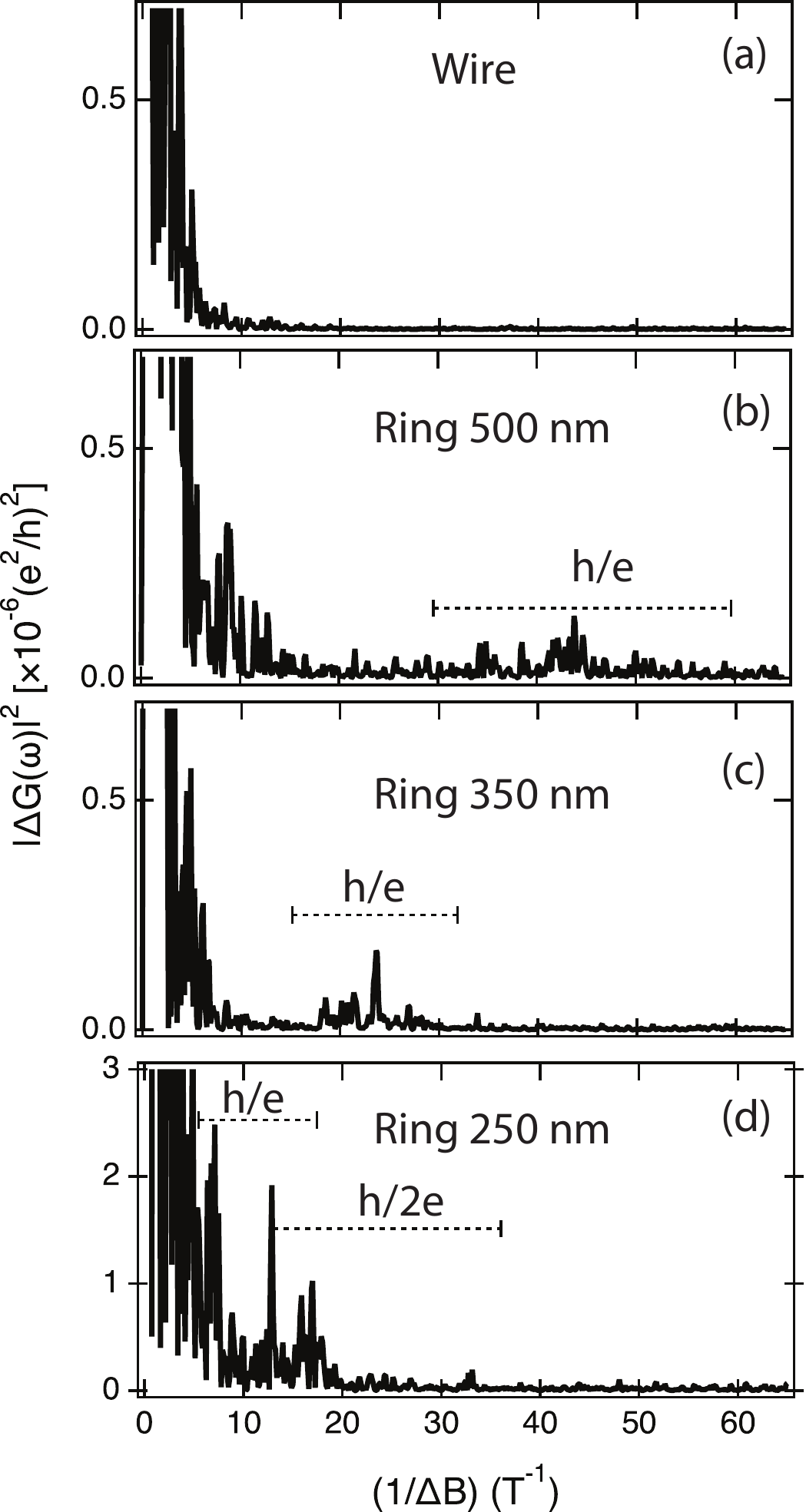}
  \caption{Fourier transform of the conductance measured at 0.1 K for a (a) Pt wire and rings with diameters: (b) 500 nm, (c) 350 nm, and (d) 250 nm.}
  \label{fig3}
\end{figure}
\begin{table}[h]
\centering
\begin{tabular}{|l|c|c|c|c|}
\hline
Sample & $d_i$ (nm) & $d_o$ (nm) & Half perimeter (nm)& $f_{h/e}$ (T$^{-1}$) \\
\hline
Ring & 180 & 310 & 393 & 6.2 - 18.3 \\
\hline
Ring & 280 & 410 & 550 &14.9 - 31.9 \\
\hline
Ring & 390 & 560 & 785 & 28.9 - 59.6 \\
\hline
\end{tabular}
\caption{\label{tab:1} Geometrical parameters of the ring samples and estimation of the expected frequency of the AB oscillations.}
\end{table}\\
\indent
To further confirm that the higher frequency peaks in the Fourier spectrum originate from the AB oscillations, we measured the magnetoresistance of other Pt rings with increasing diameters and a Pt nanowire as a control sample. The nanowire sample should have no AB oscillations, while the magnetic field periodicity for the AB ($h/e$) oscillations for the ring samples should be inversely proportional to the area enclosed by the loop of the ring. The expected periodicity of the magnetoconductance oscillations due to interference in the ring structure is given by the expression: $\Delta B_{h/e} = \phi_0/A$, where $A=\pi (d/2)^2$ is the area enclosed by the ring. Therefore, with larger ring diameters an increasing frequency for the AB oscillations is expected ($f_{h/e} = 1/\Delta B_{h/e}$) as shown in Table \ref{tab:1}, where we estimated the frequency range for the AB oscillations for each ring considering the measured inner ($d_i$) and outer ($d_o$) diameters from the SEM images (Fig. \ref{fig1}). Figure \ref{fig3} shows the Fourier spectra obtained from the conductance measurements of the different samples. We can clearly observe that the frequency of the AB peaks in the Fourier spectrum gradually shifts to larger values with increasing ring diameters (Figs. \ref{fig3}d to \ref{fig3}b). Moreover, in the case of the nanowire sample there is no AB oscillation (Fig. \ref{fig3}a), as expected. The dependence of the Fourier peak position on the sample dimension is a strong evidence of the presence of the AB oscillations in our Pt samples. It is worth noting that even in a Pt ring with a diameter of 500 nm the AB-peaks in the Fourier spectrum can still be observed, despite the half-perimeter length ($\sim$ 785 nm) being considerably larger than the phase coherence length previously estimated. Moreover, the observation of $h/e$ oscillations in the 500 nm ring, suggests that in the 250 nm ring we can possibly observe $h/2e$ oscillations corresponding to interference between electrons after performing a full loop of the ring (see Fig. \ref{fig3}d).\\
\indent
By application of an increasing bias current a larger spin accumulation can be induced by the spin Hall effect of Pt, to investigate its possible effect on the conductance fluctuations, we performed bias dependent measurements at the lowest temperature. The spin accumulation induces a shift between the electrochemical potentials for up-spin and down-spin electrons,\cite{Takahashi2008, Maekawa2013} which can have an impact on the observed interference pattern\cite{Lee1987}. Figure \ref{fig4} shows the magnetoresistance of the 250 nm ring measured at different bias currents, the sample shows a clear reduction of the resistance due to Joule heating (see Supplementary) which reduces the amplitude of the oscillations. Despite the effect of sample heating, we can identify some subtle changes in the periodicity of the observed pattern of conductance fluctuations, particularly at higher fields (B $>$ 1 T) (see highlighted area in Fig. \ref{fig4}). This variation could be related to the induced shift in the electrochemical potentials by the spin accumulation affecting the measured conductance fluctuation pattern. However, in our experiment it is not possible to perform a quantitative discussion, due to the difficulty to disentangle the contributions from sample heating and spin accumulation induced effects. Further studies using magnetic insulators as a substrate or in very thin films including a gate electrode\cite{Niita1999} can help to clarify the role of the spin accumulation on the quantum interference effects of conduction electrons.
\begin{figure}[htp!] \center
    \includegraphics[width=8.5cm,keepaspectratio]{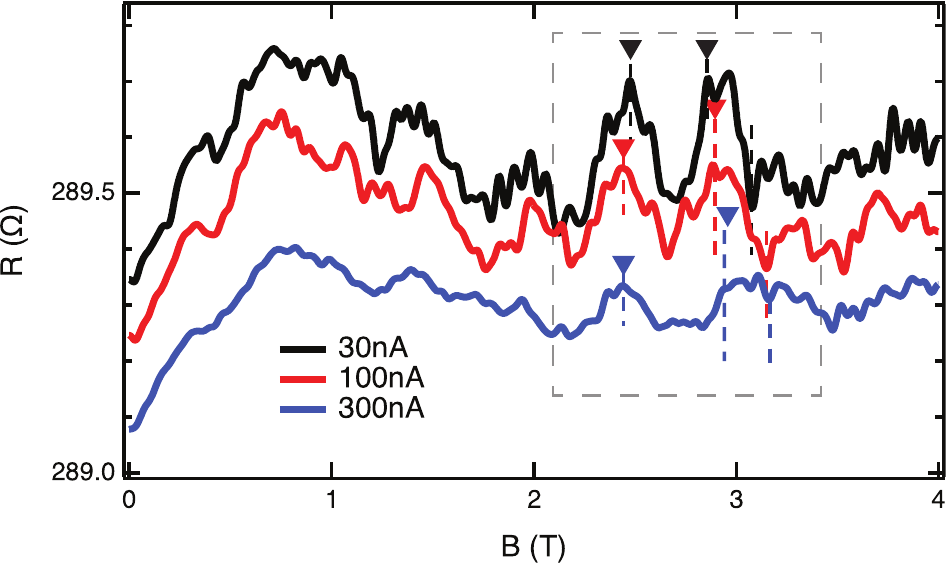}
  \caption{(Color online) Comparison of the magnetoresistance measured at the lowest temperature for 3 different bias currents: 30 nA (black line), 100 nA (red line) and 300 nA (blue line), for the Pt ring with diameter of 250 nm. High frequency components [$(\Delta \mu_0H)^{-1} >$ 18 T] were filtered out to reduce noise without losing information from UCF and AB magnetoresistance fluctuations.}
  \label{fig4}
\end{figure}\\
\indent
In summary, we have investigated the magnetoconductance of Pt rings with sub-100 nm lateral dimensions and at sub-K temperatures. We have observed clear conductance fluctuations, both periodic (AB) and aperiodic (UCF), due to interference of the electron wavefunction in metallic nanostructures. The oscillations can be observed despite the short coherence length of Pt due to the strong spin-orbit interaction in the system. We also explore their bias current dependence and observe a possible modification of the interference pattern, however further studies are needed to clarify the role of the SHE-induced spin accumulation on the quantum interference effects of the magnetoconductance.\\
The data that support the findings of this study are available from the corresponding author upon reasonable request.\\
We acknowledge T. Nojima for his support with the low temperature setup and experiments. This work was supported by ERATO ``Spin Quantum Rectification Project'' (Grant No. JPMJER1402) from JST, Japan; Grant-in-Aid for Scientific Research on Innovative Area, ``Nano Spin Conversion Science'' (Grant No. JP26103005), Grant-in-Aid for Scientific Research (S) (Grant No. JP19H05600), Grant-in-Aid for Scientific Research (C) (Grant No. JP20K05297) from JSPS KAKENHI Japan and the NEC corporation. R.R. also acknowledges support from the European Commission through the project 734187-SPICOLOST (H2020-MSCA-RISE-2016), the European Union\'{}s Horizon 2020 research and innovation program through the Marie Sklodowska-Curie Actions grant agreement SPEC number 894006 and the Spanish Ministry of Science (RYC 2019-026915-I). K.O. acknowledges support from GP-Spin at Tohoku University.

\bibliographystyle{apsrev}

\bibliography{bibliography}

\clearpage



\end{document}